\documentclass[prd,showpacs,preprintnumbers,amsmath,amssymb]{revtex4}

\usepackage{graphicx}
\usepackage{dcolumn}
\usepackage{bm}

\begin{document}

\title{Schwinger mechanism in $dS_2$ and $AdS_2$ revisited}

\author{Sang Pyo Kim}
\affiliation{Department of Physics, Kunsan National University, Kunsan 573-701,
 Korea\\Institute of Astrophysics, Center for Theoretical Physics, Department of Physics, National Taiwan University,
Taipei 106, Taiwan\\Yukawa Institute for Theoretical Physics, Kyoto University, Kyoto 606-8502, Japan}
\email{sangkim@kunsan.ac.kr}

\author{W-Y. Pauchy Hwang}
\author{Tse-Chun Wang}
\affiliation{Institute of Astrophysics, Center for Theoretical Physics, Department of Physics, National Taiwan University, Taipei 106, Taiwan}
\email{wyhwang@phys.ntu.edu.tw}

\begin{abstract}
Recalculating the Bogoliubov coefficients from the solutions in Phys. Rev. D ${\bf 78}$, 103517 (2008),
we obtain the mean number of boson pairs in a uniform electric field in the global coordinates
$dS_2$ and $AdS_2$, which have the correct zero-field and zero-curvature limits, and study
the vacuum persistence at one-loop. The mean number in $AdS_2$ gives the lowest limit to the Breitenloher-Freedman bound
in the uniform electric field, and the mean numbers in $dS_2$ and $AdS_2$ satisfy the reciprocal relation
${\cal N}_{\rm dS} (R, E) {\cal N}_{\rm AdS} (R, E) = 1$ under the analytical continuation of the scalar curvature $R$.
\end{abstract}

\date{\today}
\pacs{98.80.Cq, 04.62.+v, 12.20.-m, 03.65.Pm}

\maketitle

\section{Introduction}

The gauge-gravity interaction has recently attracted attention partly because of the running gauge coupling
due to gravity \cite{Robinson-Wilczek,Toms} and partly because of charged black holes and AdS/CFT.
It is a challenging task even to compute all one-loop diagrams for the gauge-gravity
interaction. The quantum states of charged black holes have been studied mostly near the horizon and at the spatial infinity.
The complete information of quantum states in a background gauge field and/or curved spacetime
may help one understanding the vacuum polarization and vacuum persistence,
the Heisenberg-Euler and Schwinger effective action \cite{Heisenberg-Euler,Schwinger} being the most well known
in quantum electrodynamics (QED).

The quantum states of a charged particle coupled to an electromagnetic field
may be found in a special class of spacetime, which has a certain symmetry respected
by the electromagnetic field. The solutions by Kim and Page \cite{Kim-Page08} for a
charged boson in a uniform electric field in the global coordinates of
two-dimensional de Sitter (dS) and anti-de Sitter (AdS) spaces are such an exactly solvable model.
There the spacetime has constant curvature,
and the uniform electric field respects the symmetry of the spacetime.
The symmetry of the spacetime and electric field suggests that the mean number of spontaneously produced pairs should depend on
the curvature and the electric field strength. To probe the vacuum polarization and vacuum persistence
(twice of the imaginary part of the effective action), one may employ the in-out formalism \cite{DeWitt,DeWitt-book} using
the Bogoliubov coefficients  between the in-vacuum and the out-vacuum. Thus the Bogoliubov coefficients
are important in calculating the effective action as well as the Schwinger pair production.

In this paper we revisit the Schwinger pair production and the vacuum persistence under
the uniform electric field in $dS_2$ and $AdS_2$ by recalculating the Bogoliubov coefficients using the solutions in Ref.
\cite{Kim-Page08}. For that purpose, we employ the in-vacuum and the out-vacuum in the $dS$ space as defined by Mottola, in which
the vacua are the no-particle state with respect to the positive frequency solution in each asymptotic region
\cite{Mottola,BMS,JMP}.
The procedure is similar to the Sauter-type time-dependent electric field in the Minkowski spacetime \cite{KLY08}.
For the $AdS$ space we use the vacuum in the St\"{u}ckelberg-Feynman picture that resolves the Klein paradox from a tunneling barrier
\cite{Nikishov,Damour,Hansen-Ravndal},
which has an analog of the Sauter-type space-dependent electric field in the Minkowski spacetime \cite{Kim-Page06,KLY10}.
The Schwinger pair production in $dS_2$ reduces to the particle production in the global coordinates
in the zero-field limit \cite{Mottola,BMS,JMP}, and both pair productions in $dS_2$ and $AdS_2$
reduce to the Schwinger formula in the zero-curvature limit \cite{Schwinger}.
It is shown that the pair productions in Ref. \cite{Kim-Page08} are the leading terms of the exact formula, consistent with the worldline
instanton calculation and with the planar coordinates. Further, we find the lowest limit
to the Breitenlohner-Freedman bound in the electric field in $AdS_2$ and,
surprisingly, discover the reciprocal relation between the mean numbers of produced pairs in $dS_2$ and $AdS_2$.

The organization of this paper is as follows. In Sec. II we study the periodic boundary conditions of the charged bosonic solutions.
In Sec. III we recalculate the Bogoliubov coefficients in $dS_2$, find the Schwinger pair production
and discuss the zero-field and the zero-curvature limits. In Sec. IV we show that
the pair production gives the lowest limit to the Breitenlohner-Freedman bound in the electric field in $AdS_2$.
In Sec. V we discuss the vacuum persistence and find the reciprocal relation between the mean numbers of produced pairs in
$dS_2$ and $AdS_2$.

\section{Solutions with Periodic Boundary Condition}

A uniform electric field along the spatial direction in $dS_2$ and $AdS_2$ respects the symmetry of the background geometry.
Therefore the solutions of a charged boson under the time-dependent gauge $
{\bf A}_{\rm dS} =  (0, - E \sinh (Ht)/H)$ in $dS_2$ and the space-dependent gauge ${\bf A}_{\rm AdS} =  (E \sinh (Kx)/K, 0)$
in $AdS_2$ should respect this symmetry \cite{Kim-Page08}.
The charged boson in $dS_2$ has the periodic boundary condition on $S^1$ for the Fourier momentum
\begin{eqnarray}
\Phi_{\rm dS} (t,x) = \frac{e^{ i k x}}{\sqrt{\cosh(Ht)}} \phi_k (t), \quad e^{2 \pi i p} = 1, \quad (p := \frac{k}{H} = 0,\, \pm 1,\, \pm 2,\, \cdots), \label{ds per}
\end{eqnarray}
and in $AdS_2$ has the different periodicity for the energy
\begin{eqnarray}
\Phi_{\rm AdS} (t,x) = \frac{e^{- i \omega t}}{\sqrt{\cosh(Kx)}} \phi_{\omega} (x), \quad e^{2 \pi i \tilde{p}} = 1, \quad (\tilde{p} := \frac{\omega}{K} = 0,\, 1,\, 2,\, \cdots). \label{ads per}
\end{eqnarray}

Under the periodic boundary condition (\ref{ds per})
the solution (12) of Ref. \cite{Kim-Page08} in $dS_2$ is a combination of two complex conjugates
\begin{eqnarray}
\phi_k (t) =   z^{n/2} (z^*)^{n^*/2} [c_1 F (\mu, \nu ; \gamma ; z)
+ c_2 F (\mu^*, \nu^*; \gamma^* ; z^*)], \label{ds sol}
\end{eqnarray}
where $ z = (1 + i\sinh (Ht))/2$ and $z^* = 1-z$, and with $a = 1/2 - p$
\begin{eqnarray}
n = a + i \frac{Y}{\pi}, \quad
\mu =  a - i \frac{X}{\pi}, \quad
\nu = a + i \frac{X}{\pi}, \quad
\gamma = a + \frac{1}{2} + i \frac{Y}{\pi}.
\end{eqnarray}
Here $X$ and $Y$ depend only on the scalar curvature $(R_{\rm dS} = 2 H^2)$ and the electric field as
\begin{eqnarray}
X = 2 \pi \sqrt{\Bigl(\frac{qE}{R_{\rm dS}} \Bigr)^2 + \frac{m^2}{2 R_{\rm dS}}  - \frac{1}{16}}, \quad Y = 2 \pi \frac{qE}{R_{\rm dS}}.
\label{XY}
\end{eqnarray}
Similarly, the solution (26) of Ref. \cite{Kim-Page08} in $AdS_2$ can be written as
\begin{eqnarray}
\phi_{\omega} (x) =   z^{\tilde{n}/2} (z^*)^{\tilde{n}^*/2} [d_1 F (\tilde{\mu}, \tilde{\nu} ; \tilde{\gamma} ; z)
+ d_2 F (\tilde{\mu}^*, \tilde{\nu}^*; \tilde{\gamma}^* ; z^*)], \label{ads sol}
\end{eqnarray}
where $z = (1 + i\sinh (Kx))/2$ and with $\tilde{a} = 1/2 - \tilde{p}$
\begin{eqnarray}
\tilde{n}= \tilde{a} + i \frac{\tilde{Y}}{\pi}, \quad
\tilde{\mu} =  \tilde{a} - i \frac{\tilde{X}}{\pi}, \quad
\tilde{\nu} = \tilde{a} + i \frac{\tilde{X}}{\pi}, \quad
\tilde{\gamma} = \tilde{a} + \frac{1}{2} + i \frac{\tilde{Y}}{\pi}.
\end{eqnarray}
Here $\tilde{X}$ and $\tilde{Y}$ are also functions of the curvature $(R_{\rm AdS} = -2 K^2)$ and the electric field as
\begin{eqnarray}
\tilde{X} = 2 \pi \sqrt{\Bigl(\frac{qE}{R_{\rm AdS}} \Bigr)^2 + \frac{m^2}{2 R_{\rm AdS}} - \frac{1}{16}}, \quad \tilde{Y} = - 2\pi \frac{qE}{R_{\rm AdS}}. \label{t-XY}
\end{eqnarray}

\section{Bogoliubov Coefficients in $dS_2$}

For the Bogoliubov transformation in a dS space we use the in-vacuum in the remote past and the out-vacuum in the remote future
in Refs. \cite{Mottola,BMS,JMP}. By appropriately choosing $c_1$ and $c_2$ in Eq. (\ref{ds sol}),
the positive frequency solution in the region $t \ll - 1/H $ can be expressed
in terms of the positive and negative frequency solutions in the region $t \gg 1/H$
\cite{DeWitt} (see also Appendix B of \cite{Kim11})
\begin{eqnarray}
\phi_{k, {\rm in}}^{(+)} (t) = \alpha_k \phi_{k, {\rm out}}^{(+)} (t) + \beta^*_k \phi_{k, {\rm out}}^{(-)} (t),
\end{eqnarray}
where $\phi_{k, {\rm in}}^{(\pm)}$ in $t \ll -1/H$ and $\phi_{k, {\rm out}}^{(\pm)}$ in $t \gg 1/H$
are given by
\begin{eqnarray}
\phi_{k, {\rm in}}^{(\pm)} (t) = \frac{e^{\mp i XH t/\pi}}{\sqrt{2 XH/\pi}}, \quad
\phi_{k, {\rm out}}^{(\pm)} (t) = \frac{e^{\mp i XH t/\pi}}{\sqrt{2 XH/\pi}}. \label{ds in}
\end{eqnarray}
Using the connection formula for the hypergeometric function \cite{Ab-St}
from $z = - i \infty \, (t = - \infty)$ to
$z = i \infty \, (t = \infty)$, which is valid for the Riemann sheet of ${\rm arg} (-z) < \pi$, we find Bogoliubov coefficients
\begin{eqnarray}
\alpha_k = \Bigl[2^{4 i \frac{X}{\pi}} e^{i \pi a} \frac{e^{-Y} (1+ e^{2X})}{1 - e^{2X} \frac{|B|^2}{|A|^2}}\Bigr]
\frac{B}{A}, \quad \beta_k = e^{-i\pi a} e^{X-Y} \Bigl[ \frac{1 + \frac{|B|^2}{|A|^2}}{1 - e^{2X} \frac{|B|^2}{|A|^2}}
\Bigr], \label{ds bog}
\end{eqnarray}
where
\begin{eqnarray}
A = \frac{\Gamma (\gamma) \Gamma(\nu - \mu)}{\Gamma (\nu) \Gamma (\gamma-\mu)}, \quad
B = \frac{\Gamma (\gamma) \Gamma(\mu - \nu)}{\Gamma (\mu) \Gamma (\gamma-\nu)}.
\end{eqnarray}

The periodic boundary condition (\ref{ds per}) leads to the vacuum persistence amplitude and the
Schwinger pair production
\begin{eqnarray}
|\alpha_k|^2 = \frac{\cosh (X+Y) \cosh (X-Y)}{(\sinh X)^2}, \quad |\beta_k|^2 = \Bigl(\frac{\cosh Y}{\sinh X} \Bigr)^2. \label{ds rate}
\end{eqnarray}
These are independent of the momentum $k$,
depend only on the scalar curvature and the electric field strength, and thus
respect the symmetry of $dS_2$. Also the Bogoliubov relation $(|\alpha_k|^2 - |\beta_k|^2 =1)$ is satisfied.
The overall factor for the mean number of produced pairs (${\cal N} = |\beta|^2$)
\begin{eqnarray}
{\cal N}_{\rm dS} = e^{-2 (X-Y)} \Bigl(\frac{1 + e^{-2Y}}{1-e^{-2X}} \Bigr)^2, \label{ds lead}
\end{eqnarray}
is the leading term given in Eqs. (20) and (32) of Ref. \cite{Kim-Page08}.
In the zero-curvature limit $(R_{\rm dS} = 0)$, Eq. (\ref{ds rate}) reduces to the Schwinger formula in the Minkowski spacetime and in the zero-electric field limit $(E = 0)$, it recovers the particle production in the global coordinates of $dS_2$ \cite{Mottola,BMS,JMP,Kim10b}.
The pair-production rate is obtained by multiplying (\ref{ds lead}) by the number of states $qE/(2\pi)$ per unit volume and per unit time.

\section{Bogoliubov Coefficients in $AdS_2$}

The charged field in the AdS space can be quantized by the same method as QED in a space-dependent gauge in which the Klein paradox from
a tunneling barrier should be resolved \cite{Nikishov,Damour,Hansen-Ravndal}. In the St\"{u}ckelberg-Feynman picture,
pair annihilation is the process in which
a particle $\phi_{\omega, {\rm in}}^{(+)}$ from $x = - \infty$ interacts with the electric field and/or the spacetime curvature
and then travels backward in time as an antiparticle $\phi_{\omega, {\rm in}}^{(-)}$ to $x = \infty$. In contrast,
pair production is another process in which
an antiparticle $\phi_{\omega, {\rm out}}^{(-)}$ travels backward in time from $x = \infty$ and then
travels forward in time as a particle $\phi_{\omega, {\rm out}}^{(+)}$ to $x = - \infty$.
Again by appropriately choosing $d_1$ and $d_2$ in Eq. (\ref{ads sol}), the in-vacuum solution at $x \ll - 1/K$ with respect to the group velocity
can be expressed as (see, for instance, Ref. \cite{Kim11})
\begin{eqnarray}
e^{- i \omega t} \phi_{\omega, {\rm in}}^{(+)} (x) = \frac{\tilde{\alpha}_{\omega}}{\tilde{\beta}_{\omega}}
e^{i \omega (-t)} \phi_{\omega, {\rm in}}^{(-)} (x) + \frac{1}{\tilde{\beta}^*_{\omega}} e^{i \omega (-t)}
\phi_{\omega, {\rm out}}^{(-)} (x), \label{ads bog tr}
\end{eqnarray}
where $\phi_{\omega, {\rm in}}^{(+)}$ and $\phi_{\omega, {\rm out}}^{(+)}$ in $x \ll -1/K$
and $\phi_{\omega, {\rm in}}^{(-)}$ and $\phi_{\omega, {\rm out}}^{(-)}$ in $x \gg 1/K$
are
\begin{eqnarray}
\phi_{\omega, {\rm in}}^{(\pm)} (x) = \frac{e^{i \tilde{X}K x/\pi}}{\sqrt{2 \tilde{X}K/\pi}},
\quad \phi_{\omega, {\rm out}}^{(\pm)} (x) = \frac{e^{-i \tilde{X}K x/\pi}}{\sqrt{2 \tilde{X}K/\pi}}. \label{ads min}
\end{eqnarray}

Then the Bogoliubov coefficients are given by
\begin{eqnarray}
\tilde{\alpha}_{\omega} = \Bigl[ 2^{4 i \frac{\tilde{X}}{\pi}} e^{- 2 i \pi \tilde{a}} \frac{e^{\tilde{X}} (1+ e^{-2\tilde{X}})}{1 + \frac{|\tilde{B}|^2}{|\tilde{A}|^2} } \Bigr] \frac{\tilde{B}}{\tilde{A}}, \quad
\tilde{\beta}_{\omega} = e^{- i\pi \tilde{a}} e^{\tilde{X}-\tilde{Y}} \Bigl[\frac{1 - e^{-2\tilde{X}} \frac{|\tilde{B}|^2}{|\tilde{A}|^2}
}{1 + \frac{|\tilde{B}|^2}{|\tilde{A}|^2} } \Bigr], \label{ads bog}
\end{eqnarray}
where
\begin{eqnarray}
\tilde{A} = \frac{\Gamma (\tilde{\gamma}) \Gamma(\tilde{\nu} - \tilde{\mu})}{\Gamma (\tilde{\nu}) \Gamma (\tilde{\gamma}-\tilde{\mu})}, \quad
\tilde{B} = \frac{\Gamma (\tilde{\gamma}) \Gamma(\tilde{\mu} - \tilde{\nu})}{\Gamma (\tilde{\mu}) \Gamma (\tilde{\gamma}-\tilde{\nu})}.
\end{eqnarray}
The vacuum persistence amplitude and the Schwinger pair production
\begin{eqnarray}
|\tilde{\alpha}_{\omega}|^2 = \frac{\cosh (\tilde{X}+\tilde{Y}) \cosh(\tilde{X}-\tilde{Y})}{(\cosh \tilde{Y})^2}, \quad |\tilde{\beta}_{\omega}|^2 = \Bigl(\frac{\sinh \tilde{X}}{\cosh \tilde{Y}} \Bigr)^2, \label{ads rate}
\end{eqnarray}
depend only the scalar curvature and the electric field, and the Bogoliubov relation $(|\tilde{\alpha}_{\omega}|^2 - |\tilde{\beta}_{\omega}|^2 =1)$
holds. Noting $\tilde{Y} > \tilde{X}$, the mean number of produced pairs can be written as
\begin{eqnarray}
{\cal N}_{\rm AdS} = |\beta_{\omega}|^2 = e^{-2 (\tilde{Y} - \tilde{X})} \Bigl(\frac{1- e^{-2\tilde{X}}}{1+ e^{-2\tilde{Y}}} \Bigr)^2.
\label{ads lead}
\end{eqnarray}
The overall factor is the leading term in Eqs. (31) and (32) of Ref. \cite{Kim-Page08}.
In the global coordinates the pair production can completely halt when $\tilde{X} = 0$,
which gives the lowest limit (equality) to the Breitenlohner-Freedman bound \cite{Breitenlohner-Freedman}
in the presence of a uniform electric field in the Poincar\'{e} patch by Pioline and Troost \cite{Pioline-Troost}
\begin{eqnarray}
m^2 \geq \Bigl(\frac{qE}{K} \Bigr)^2  - \frac{K^2}{4}. \label{BF boun}
\end{eqnarray}
In the zero-curvature limit the pair production reduces to the Schwinger formula but the zero-curvature limit
does not exist due the bound (\ref{BF boun}).

\begin{figure}[t]
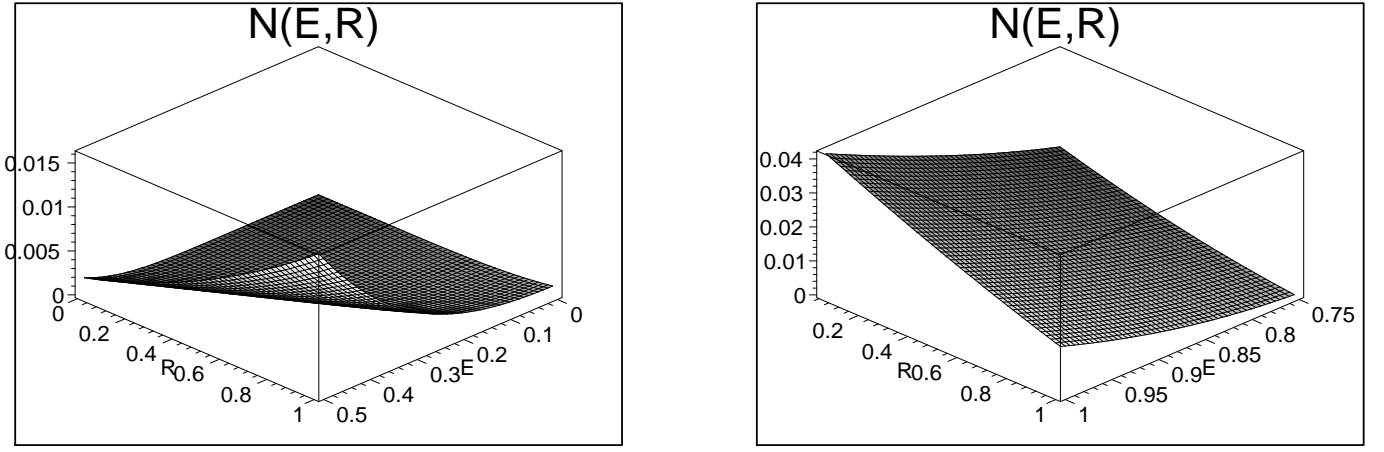

{\includegraphics[width=0.45\linewidth,height=0.25\textheight]{ds-PR1.eps} }\hfill
{\includegraphics[width=0.45\linewidth,height=0.25\textheight
]{ads-PR1.eps}}
\caption{In weak electric fields, the pair production rate in $dS_2$ as a function of the rescaled variables $E: = E/(qm^2)$ and $R:= R_{\rm dS}/m^2$ [left panel] and in $AdS_2$ beyond the Breitenlohner-Freedman bound
  as a function of the rescaled variables $E:= E/(qm^2)$ and $R:= - R_{\rm AdS}/m^2$ [right panel].} \label{fig1}
\end{figure}

A closer inspection of Eqs. (\ref{XY}) and (\ref{t-XY}) shows that the ratio of the Maxwell scalar to the scalar curvature,
$(qE/m^2)^2/(R/m^2)$,
determines the dominance of the Schwinger mechanism or the scalar curvature for pair production.
In the weak field regime for  $dS_2$,
though both the electric field and the scalar curvature enhance pair production,
the curvature drastically enhances pair production
as shown on the the left panel of Fig. \ref{fig1}. On the other hand,
the confinement in $AdS_2$ reflected in the negative scalar curvature suppresses pair production as shown
on the right panel of Fig. \ref{fig1}.
In the strong field regime the pair productions increase as the electric
field increases but they are almost independent of small scalar curvature and thus do not distinguish
$dS_2$ from $AdS_2$.

\section{Vacuum Persistence and Reciprocal Relation}

The bosonic vacuum persistence ($2 \, {\rm Im} (W) = \sum \ln (|\alpha|^2)$) at one-loop
is for $dS_2$
\begin{eqnarray}
2 \, {\rm Im} (W_{\rm dS}) = \sum_{k} \ln \Bigl( \frac{\cosh (X+Y) \cosh (X-Y)}{(\sinh X)^2} \Bigr),
\end{eqnarray}
and for $AdS_2$
\begin{eqnarray}
2 \, {\rm Im} (W_{\rm AdS}) = \sum_{\omega} \ln \Bigl(\frac{\cosh (\tilde{X}+\tilde{Y}) \cosh(\tilde{X}-\tilde{Y})}{(\cosh \tilde{Y})^2} \Bigr).
\end{eqnarray}
The factorization in hyperbolic functions suggests that the Heisenberg-Euler and Schwinger effective action \cite{Schwinger}
may have some analogs in $dS_2$ and $AdS_2$.
A more interesting interpretation occurs when the scalar curvature is analytically continued to different regimes:
the negative curvature for $dS_2$ and the positive curvature for $AdS_2$ all within the bound
\begin{eqnarray}
\Bigl(\frac{qE}{R} \Bigr)^2 + \frac{m^2}{2 R}  \geq \frac{1}{16}. \label{an con}
\end{eqnarray}
The bound (\ref{an con}) is the condition for pair production both in $dS_2$ and
$AdS_2$ and forms the set complementary of the Breitenlohner-Freedman bound for $AdS_2$.
For instance, the analytical continuation
\begin{eqnarray}
t = i \zeta, \quad x = i \tau, \quad H = i {\cal H}, \quad K = i {\cal K}
\end{eqnarray}
connects $dS_2$ to $AdS_2$ through $R = 0$ and vice versa in the superspace of constant scalar curvatures.

Remarkably, the pair productions (\ref{ds rate}) and (\ref{ads rate}) under the analytical continuation with (\ref{an con})
have the reciprocal relation
\begin{eqnarray}
{\cal N}_{\rm dS} (R, E) {\cal N}_{\rm AdS} (R, E) = 1, \label{rec}
\end{eqnarray}
where $R$ denotes $R_{\rm dS}$ or $R_{\rm AdS}$ but not both of them simultaneously.
The reciprocal relation (\ref{rec}) and the Bogoliubov relations
lead to the vacuum persistence in another form
\begin{eqnarray}
2 \, {\rm Im} (W) = - \sum_{k} \ln ( 1 - {\cal N}^{\rm sp}), \label{vac inv}
\end{eqnarray}
where
\begin{eqnarray}
{\cal N}_{\rm dS}^{\rm sp} := \frac{1}{|\alpha_{\rm AdS}|^2}, \quad {\cal N}_{\rm AdS}^{\rm sp} := \frac{1}{|\alpha_{\rm dS}|^2}.
\end{eqnarray}
Note that the vacuum persistence (\ref{vac inv}) has the form for charged fermions, which is reminiscent of the spin-statistics inversion of Schwinger
pair production in QED \cite{Stephens,Hwang-Kim09}, for instance, the bosonic vacuum persistence
\begin{eqnarray}
2 \, {\rm Im} (W) = - \sum_{k} \ln ( 1 - {\cal N}_{\rm FD}), \label{sp inv}
\end{eqnarray}
is given by the Fermi-Dirac distribution  ${\cal N}_{\rm FD}$ and vice versa.

\section{Conclusion}

We have recalculated the Bogoliubov coefficients for the solutions in Ref. \cite{Kim-Page08}
for a charged boson in a uniform electric field in $dS_2$ and $AdS_2$. The coefficients are
important not only in calculating the Schwinger pair production but also
finding the effective action and the vacuum persistence in the in-out formalism \cite{DeWitt-book}.
In QED in the Minkowski spacetime the effective actions and vacuum persistence have been found in a time-dependent electric field \cite{KLY08}
and in a space-dependent electric field of Sauter-type \cite{KLY10}. The vacuum persistence in this paper is an important tool
to check the consistency of QED effective actions both in the uniform electric field and in $dS$ and $AdS$, which will be addressed
in the future.

\acknowledgments
S.~P.~K. thanks Don N. Page for valuable comments and Christian Schubert for useful discussions. He
also thanks Misao Sasaki for the warm hospitality at Yukawa Institute for Theoretical Physics, Kyoto University.
The work of S.~P.~K. was supported by National Science Council Grant (NSC 100-2811-M-002-012)
and the work of W-Y.~P.~H.~was supported in part by National Science Council Project (NSC 99-2112-M-002-009-MY3).


\begin{references}

\bibitem{Robinson-Wilczek} S.~P.~Robinson and F.~Wilczek, Phys.\ Rev.\ Lett.\ {\bf 96}, 231601 (2006).

\bibitem{Toms} D.~J.~Toms, Nature {\bf 468}, 56 (2010).

\bibitem{Heisenberg-Euler} W.~Heisenberg and H.~Euler, Z. Phys. {\bf 98}, 714 (1936)  [physics/0605038].

\bibitem{Schwinger} J.~Schwinger, Phys. Rev. {\bf 82}, 664 (1951).

\bibitem{Kim-Page08} S.~P.~Kim and D.~N.~Page, Phys. Rev. D {\bf 78}, 103517 (2008).

\bibitem{DeWitt} B.~S.~DeWitt, Phys.\ Rep.\ {\bf 19}, 295 (1975).

\bibitem{DeWitt-book} B.~S.~DeWitt, {\it The Global Approach to Quantum Field Theory}
(Oxford University Press, New York, 2003) Vol. 1 and Vol. 2.

\bibitem{Mottola} E.~Mottola, Phys. Rev. D {\bf 31}, 754 (1985).

\bibitem{BMS} R.~Bousso, A.~Maloney, and A.~Strominger, Phys. Rev. D {\bf 65},
104039 (2002).

\bibitem{JMP} E.~Joung, J.~Mourad and R.~Parentani, J. High Energy Phys. 08 (2006) 082.

\bibitem{KLY08} S.~P.~Kim, H.~K.~Lee, and Y.~Yoon,  Phys. Rev. D {\bf 78},  105013 (2008).

\bibitem{Nikishov} A.~I.~Nikishov, Nucl. Phys. B {\bf 21},  346 (1970).

\bibitem{Damour} T.~Damour, in {\it Proceedings of the First Marcel Grossmann Meeting on General Relativity}, edited by R.~Ruffini (North-Holland, Amsterdam, 1977) pp. 459.

\bibitem{Hansen-Ravndal} A.~Hansen and F.~Ravndal, Phys. Scr. {\bf 23}, 1036 (1981).

\bibitem{Kim-Page06} S.~P.~Kim and D.~N.~Page, Phys.\ Rev.\ D {\bf 73}, 065020 (2006).

\bibitem{KLY10} S.~P.~Kim, H.~K.~Lee, and Y.~Yoon, Phys. Rev. D {\bf 82}, 025015 (2010).

\bibitem{Kim11} S.~P.~Kim, ``QED Effective Actions in Space-Dependent Gauge and Electromagnetic Duality,'' arXiv:1105.4382 [hep-th].

\bibitem{Ab-St} M.~Abramowitz and  I.~Stegun, {\it Handbook of Mathematical
Functions} (Dover Publications Inc., New York, 1964), the asymptotic form 15.3.7.

\bibitem{Kim10b} S.~P.~Kim, J. High Energy Phys. 09 (2010) 054.

\bibitem{Breitenlohner-Freedman} P. Breitenlohner and D.~Z.~Freedman, Ann. Phys. (N.Y.) {\bf 144}, 249 (1982).

\bibitem{Pioline-Troost} B.~Pioline and J.~Troost, J. High Energy Phys. 03 (2005) 043.

\bibitem{Stephens} C.~R.~Stephens, Ann. Phys. (N.Y.) {\bf 193}, 255 (1989).

\bibitem{Hwang-Kim09} W-Y.~Pauchy Hwang and S.~P.~Kim, Phys. Rev. D {\bf 80}, 065004 (2009).

\end{references}
\end{document}